\documentstyle[graphicx,latexsym]{aipproc}

\newcommand{\ltsim}{\lower.5ex\hbox{$\; \buildrel < \over \sim \;$}}
\newcommand{\gtsim}{\lower.5ex\hbox{$\; \buildrel > \over \sim \;$}}

\newcommand{\aprx}	{\mbox{$\sim$}}

\newcommand{\etal}              {{\it et~al.\ }}

\newdimen\digitwidth
{\catcode`?=\active
        \gdef\setmathspace   {
            \setbox0=\hbox{\rm0}
            \digitwidth=\wd0
            \catcode`?=\active
            \def?{\kern\digitwidth}
        }
}

\def\sl2{{Spacelab~2}}

\def\heaoone{{\em HEAO-1}}
\def\heao3{{\it HEAO 3}}

\newcommand{\rxte}{{\em RXTE}}
\newcommand{\sax}{{{\em Beppo}SAX}}

\newcommand{\hxt}{HEXTE}

\def\a0{\mbox{A~0535+26}}
\def\grs1915{\mbox{GRS~1915+105}}
\def\GX354{\mbox{GX354+0}}

\def\j1744{\mbox{GRO~J1744-28}}

\def\degree{\hbox{$^\circ$}}
\def\cmsq{{$\rm cm^2$}}

\newcommand{\xray}	{\mbox{X--ray}}

\newcommand{\srcnm}{{\mbox{4U~0115+63}}}

\begin{document}

\title{Multiple Cyclotron Lines in the Spectrum of \srcnm}

\author{W.A. Heindl$^*$, W. Coburn$^*$, D.E. Gruber$^*$,
M. Pelling$^*$, R.E. Rothschild$^*$
J. Wilms$^{\dagger}$,
K. Pottschmidt$^{\dagger}$, and R. Staubert$^{\dagger}$}

\address{$^*$Center for Astrophysics and Space Sciences, University of
California San Diego,\\La Jolla, CA, 92093, USA\\$^{\dagger}$Institut
f\"{u}r Astronomie und Astrophysik -- Astronomie,\\Waldh\"{a}user
Str. 64, D-72076 T\"{u}bingen, Germany}

\maketitle

\begin{abstract}
We report phase resolved spectroscopy of the transient accreting
pulsar, \srcnm.  For the first time, more than two cylotron resonance
scattering features are detected in the spectrum of an \xray\ pulsar.
The shape of the fundamental line appears to be complex, and this is
in agreement with predictions of Monte-Carlo models.  As in other
pulsars, the line energies and optical depths are strong functions of
pulse phase.  One possible model for this is an offset of the dipole
of the neutron star magnetic field.

\end{abstract}

\section{Introduction}

\srcnm\ is a transient accreting \xray\ pulsar in
an eccentric 24 day orbit \cite{bil97} with an O9e star \cite{ung98}.
A cyclotron resonance scattering feature (CRSF) was first
noted near 20\,keV by Wheaton, \etal\ (1979) \nocite{whe79} with the
UCSD/MIT hard \xray\ (10-100\,keV) experiment aboard \heaoone. White,
Swank \& Holt (1983) \nocite{wsh83} analyzed concurrent data from the lower energy
(2-50\,keV) \heaoone/A2 experiment and found an additional feature at
\aprx12\,keV, making \srcnm\ the first pulsar with two cyclotron line
harmonics.

We discuss here phase-resolved spectra derived from an observation
of the 1999 March outburst \cite{wil99,hei99} obtained with the
\emph{Rossi X-Ray Timing Explorer} (\rxte).  First results of this
work are detailed in Heindl \etal\ (1999)\nocite{hei99}.  \sax\ has
also made detailed observations of this outburst \cite{San99a,San99b}.

\section{Observations and Analysis}

Observations were made with the Proportional
Counter Array (PCA) \cite{jah96} and High Energy X-ray Timing
Experiment (\hxt) \cite{rot98} on board \rxte.  The PCA is a set of 5
Xenon proportional counters sensitive in the
energy range 2--60\,keV with a total effective area of \aprx
7000\,$\rm cm^2$.  \hxt\ consists of two arrays of 4 NaI(Tl)/CsI(Na)
phoswich scintillation counters (15-250\,keV) totaling \aprx 1600
\cmsq. The \hxt\ alternates between source and background
fields in order to measure the background.  The PCA and \hxt\ fields
of view are collimated to the same 1\degree\ full width half
maximum (FWHM) region.

We performed four long pointings (duration \aprx15-35\,ks to search
for CRSFs.  The second observation, on 1999 March 11.87-12.32, spanned
periastron passage at March 11.95 (\cite{bil97}) and preceded the
outburst maximum by about 2\,days.  The results presented here are
from this observation.
\begin{figure}
\centerline{\includegraphics[height=3in,bb=106 488 363 685]{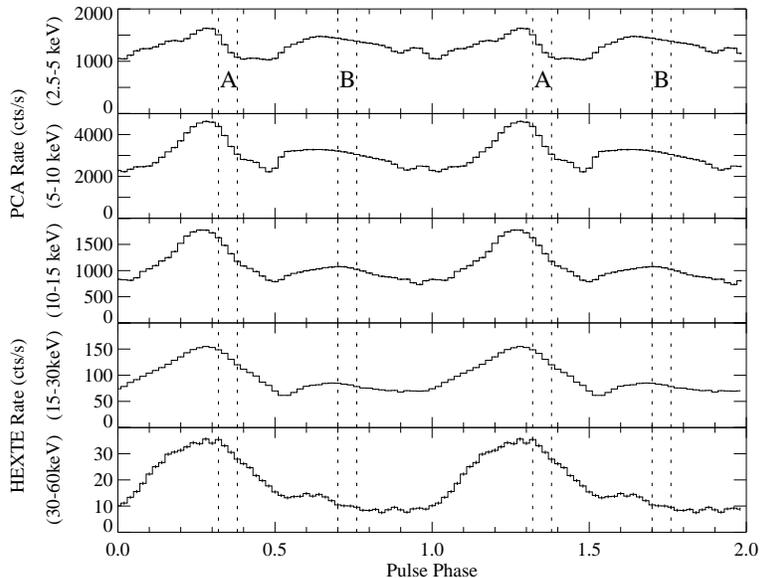}}
\caption{\label{fig:flc}The folded light curve (or ``pulse profile'')
of \srcnm\ in five energy bands.  We report detailed spectral
analyses in phase ranges A and B which are indicated by the dashed
lines.}
\end{figure}

The spectrum of \srcnm\ varies significantly with neutron star
rotation phase, making fits to the average spectrum difficult to
interpret. To avoid this problem, we accumulated spectra as a function
of pulse phase. Figure~\ref{fig:flc} shows the folded light curve in 5
energy bands. The pulse is double peaked at low energies, but the
second peak nearly disappears at high energy. Two phase bands, ``A''
and ``B'', which we selected for detailed analysis, are indicated. Our
spectral fitting process consists of fitting the joint PCA/HEXTE
spectra to various heuristic models (see Kreykenbohm, \etal\
1998\nocite{kre98}) which have been successful in fitting pulsars with
no cyclotron lines. When none of these models can adequately describe
the spectrum, and line-like residuals are present, we add Gaussian
absorption lines to the spectrum as required. For a detailed
description of our analysis technique, see Heindl, \etal\
(1999)\nocite{hei99}.  
\begin{figure}
\centerline{\includegraphics[width=7cm,bb=133 157 529 518]{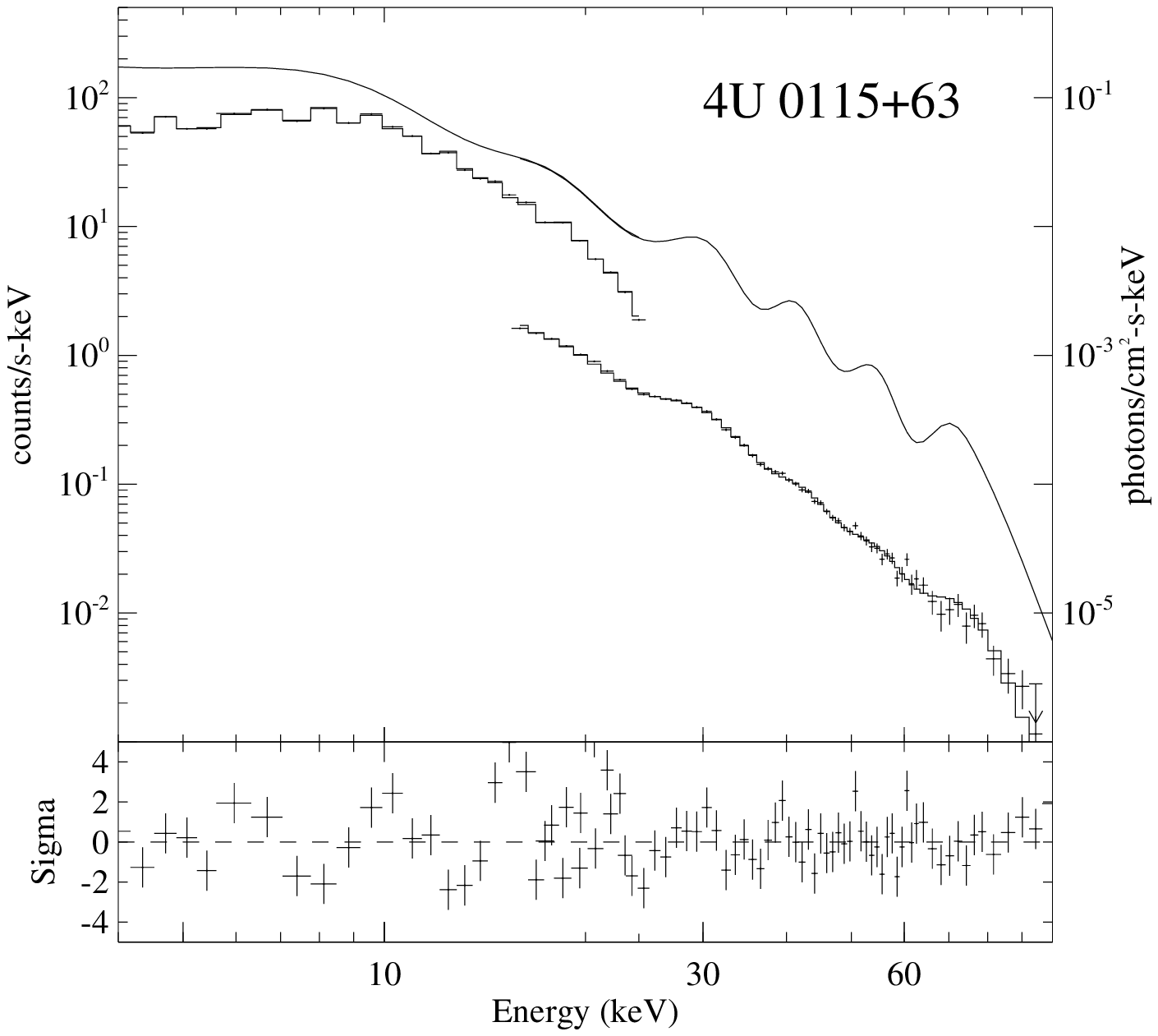}
\hspace{0.25cm}
\includegraphics[width=7cm,bb=121 133 529 518]{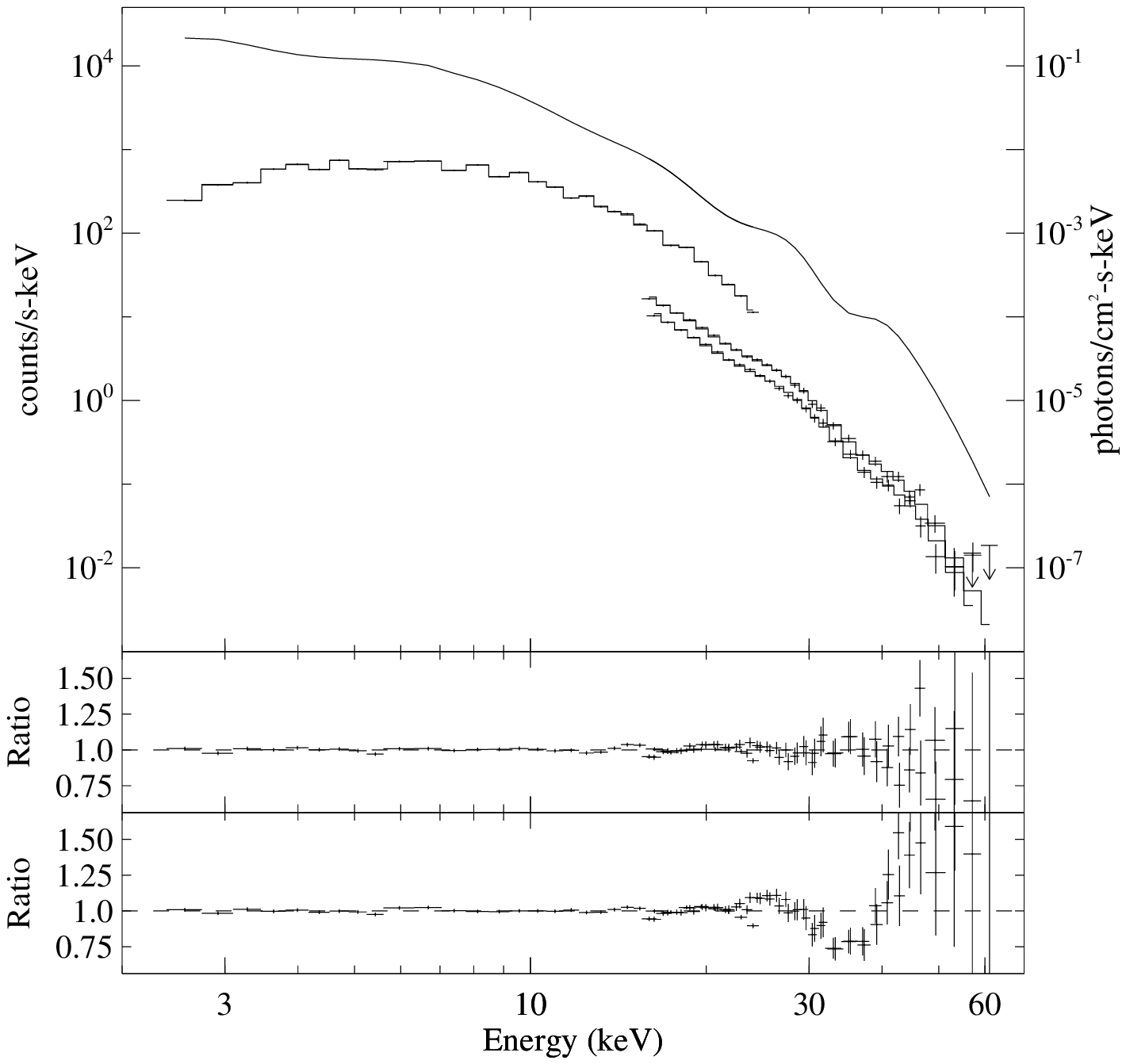}}
\caption{\label{fig:spectra} Best fit spectra to (left) phase A and
(right) phase B. The top panel shows the inferred incident spectrum
(smooth curve), the measured count spectra (data points), and the
model count spectra (histograms).  The bottom panels show residuals.
For phase B, residuals fits with three (middle) and two (bottom)
cyclotron lines. }
\end{figure}

\begin{figure}
\centerline{\includegraphics[height=3.in,bb=127 128 535 534]{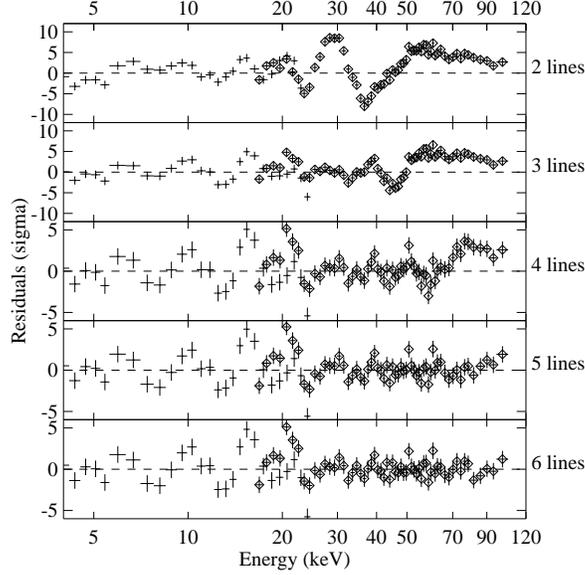}}
\caption{\label{fig:delchi}Residuals to fits to the phase A spectrum
   with different numbers of cyclotron line harmonics.  Five lines are
   required to adequately fit the data.}
\end{figure}

Figure~\ref{fig:spectra} shows the resulting best fit spectra for
phases A and B.  Both fits have a Fermi-Dirac cutoff powerlaw
continuum, a low energy excess in the form of a black body with
kT\aprx0.4\,keV, \emph{no} FeK line, and multiple cyclotron lines. In
phase A, the falling edge of the main pulse, five cyclotron lines are
required, while in phase B, the fall of the second pulse, only three
lines are necessary. Figure~\ref{fig:delchi} illustrates how we
determine how many lines are present.  The five panels show the
residuals to phase A fits with increasing numbers of lines.  With
fewer than five lines, significant residuals are present.  These
residuals follow the pattern of a dip at the energy of the first
missing line and a gross underprediction of the continuum at high
energies.  The fit continuum tends to conform to the low side of the
missing line, because the statistics are rapidly decreasing with
energy.  The too steeply falling continuum can then never recover the
high energy data.  In phase A, adding five harmonics significantly improves
the fit, while a sixth is not statistically required.

\section{Results and Discussion}
Table~\ref{tab:fits} lists the best fit cyclotron line parameters. Two
interesting phenomena are apparent from these results.  First, within
the individual phases, the lines are not harmonically spaced. And,
second, the line energies vary significantly with pulse phase.

A harmonic relation (with small modifications due to relativistic
effects) between line energies is expected in simple Landau
theory. However, models of cyclotron line formation (e.g. Isenberg,
Lamb, \& Wang 1998\nocite{ise98}) predict that the fundamental line
shape can be quite complex, even having wings resembling
\emph{emission} features.  On the other hand, the higher harmonics
should have relatively simple line shapes.  Thus, it may be that our
simple Gaussian absorption does not give an accurate measure of the
fundamental line energy.  Evidence for this appears in the residuals
near 10--15\,keV (see Fig.~\ref{fig:delchi}), which are the most
significant remaining deviations in our best fit.  Furthermore the
third and higher harmonics' energies are consistent with integer
multiples of \emph{half} of the second harmonic energy.  Thus it seems
likely that \emph{half} of the second harmonic energy gives a more
accurate measure of the magnetic field.  In the case of phase A, this
is $\rm B = 1.3\times10^{12}$\,G, assuming a gravitational redshift of
0.3 to the neutron star surface.
\begin{table}
\caption{\label{tab:fits} Best fit cyclotron line parameters from the two phase
bins shown in Fig.~\ref{fig:flc}.}
\begin{minipage}{\linewidth}
\renewcommand{\thefootnote}{\thempfootnote}
\begin{center}
\begin{tabular}{lccclccc} \hline \hline
 & \multicolumn{3}{c}{Phase A} & & \multicolumn{3}{c}{Phase B}\\
\cline{2-4} \cline{6-8}
 Harmonic & Energy & Width\protect\footnote{One standard deviation of the Gaussian optical
   depth profile} & Optical Depth && Energy & Width\protect\footnotemark & Optical Depth\\
   & (keV)  & (keV) & & & (keV)  & (keV)\\ \hline
 1\protect\footnote{also called the ``fundamental''.} & $\rm 13.35^{+0.08}_{-0.06}$ & $\rm  3.29^{+0.13}_{-0.07} $ 
		& $\rm 1.17^{+0.06}_{-0.04} $
   && $\rm 12.40^{+0.65}_{-0.35}$ & $\rm  3.3^{+0.19}_{-0.4} $ 
		& $\rm 0.72^{+0.10}_{-0.17}$ \\
 2 & $\rm 23.7{\pm0.1} $ &$\rm 5.43^{+0.18}_{-0.27} $ 
		& $\rm 2.68^{+0.05}_{-0.07}$
   && $\rm 21.45^{+0.25}_{-0.38} $ &$\rm 4.5^{+0.7}_{-0.9} $ 
		& $\rm 1.24^{+0.04}_{-0.06} $\\
 3 & $\rm 36.4^{+0.4}_{-0.5} $ & $\rm 4.3^{+0.4}_{-0.6}$ 
		& $\rm 2.41^{+0.11}_{-0.13} $
   && $\rm 33.56^{+0.70}_{-0.90} $ & $\rm 3.8^{+1.5}_{-0.9}$ 
		& $\rm 1.01^{+0.13}_{-0.14} $\\ 
 4 & $\rm 47.8^{+0.4}_{-0.7} $ & $\rm 5^{+\infty}_{-1.2}$ 
		& $\rm 2.3{\pm 0.2} $
   && -- & -- & -- \\ 
 5 & $\rm 61.7{\pm 1.1} $ & $\rm 5 \emph{fixed} $ 
		& $\rm 1.8{\pm 0.3} $
   && -- & -- & -- \\ \hline\hline
\end{tabular}
\end{center}
\end{minipage}
\end{table}

\begin{figure}
\centerline{\includegraphics[width=3.5in,bb=116 82 529 522]{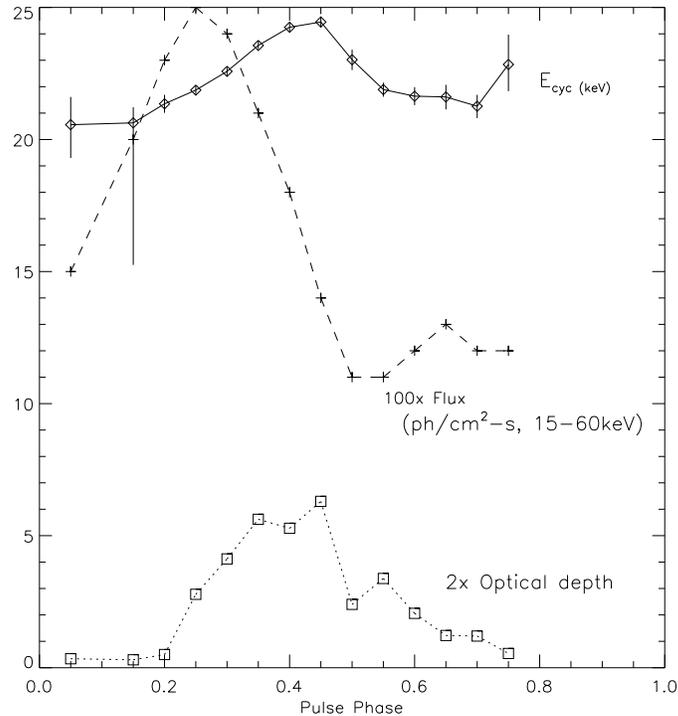}}
\caption{\label{fig:linevar}As a function of pulse phase, the energy
   (top) and depth (bottom) of the \aprx20\,keV cyclotron line.  Also
   plotted is the flux in the HEXTE band.}
\end{figure}

Figure~\ref{fig:linevar} shows the HEXTE flux and the \aprx20\,keV
line energy and optical depth as a function of pulse phase.  These
parameters were determined from the HEXTE data alone in 20 phase bins.
The second harmonic line was not required in the HEXTE data \emph{alone}
at phases greater than 0.7.  Both the line energy and optical depth
are maximal not at the pulse peak, but on the falling edge of the main
pulse.  The line energy varies by \aprx20\%.  This behavior is also
seen in Cen~X-3 \cite{hei99b}.  In Cen~X-3, Burderi et al. (2000)
\nocite{Bur00} have modeled the variation of the line energy as
due to an offset of the magnetic dipole moment from the center of the
neutron star, and this may also be the case in \srcnm.

{\bibliographystyle{amsplain}
\bibliography{proceedings}}

\providecommand{\bysame}{\leavevmode\hbox to3em{\hrulefill}\thinspace}
\begin{thebibliography}{10}

\bibitem{bil97}
L.~Bildsten et~al., ApJS \textbf{113} (1997), 367.

\bibitem{Bur00}
L.~Burderi et~al., ApJ \textbf{530} (2000), 429.

\bibitem{hei99}
W.~A. Heindl et~al., ApJ \textbf{521} (1999), L49.

\bibitem{hei99b}
W.A. Heindl and D.~Chakrabarty, 1999, to appear in MPE Report: ``Highlights in
  X-ray Astronomy in Honour of Joachim Tr{\"u}mper's 65th Birthday''.

\bibitem{jah96}
K.~Jahoda et~al., SPIE \textbf{2808} (1996), 59.

\bibitem{kre98}
I.~Kreykenbohm et~al., A\&A \textbf{341} (1998), 141.

\bibitem{rot98}
R.E. Rothschild et~al., ApJ \textbf{496} (1998), 538.

\bibitem{San99a}
A.~Santangelo et~al., 1999, in this proceedings.

\bibitem{San99b}
\bysame, ApJ \textbf{523} (1999), L85.

\bibitem{ung98}
S.J. Unger, P.~Roche, I.~Negueruela, F.A. Ringwald, C.~Lloyd, and M.J. Coe,
  A\&A \textbf{336} (1998), 960.

\bibitem{whe79}
W.~A. Wheaton et~al., Nature \textbf{282} (1979), 240.

\bibitem{wsh83}
N.E. White, J.H. Swank, and S.S. Holt, ApJ \textbf{270} (1983), 711.

\bibitem{wil99}
R.B. Wilson, B.A. Harmon, and M.H. Finger, IAU Circ. (1999), No. 7116.

\end{thebibliography}

\end{document}